\documentstyle[12pt]{article}
\begin{document}
\input epsf

\title{Finite Size Scaling of Perceptron}
\author{E.Korutcheva{\thanks{Corresponding author: fax: +34-91-398-66-97, 
e-mail: elka@fisfun.uned.es; \,\,\ 
Permanent address: G.Nadjakov Inst. Solid State
Physics, Bulgarian Academy of Sciences, 1784 Sofia, Bulgaria}}\\
Dep. F\'{\i}sica Fundamental\\Universidad Nacional de Educaci\'on a Distancia,\\ c/ Senda del Rey No 9 ,  28080 Madrid, Spain\\
and\\
N.Tonchev,\\
G.Nadjakov Institute of Solid State Physics,\\
Bulgarian Academy of Sciences,
1784 Sofia, Bulgaria}
\date{}
\maketitle

\vskip1cm
\begin{abstract}
We study the first-order phase transition in the model of a simple
perceptron with continuous weights and large, 
but finite value of the inputs. 
Making the analogy with the  usual finite-size
physical systems, we calculate the
shift and the rounding exponents near the transition point. 
In the case of a general perceptron 
with larger variety of outputs, the analysis gives only bounds for
the exponents.
\end{abstract}


keywords: neural networks, perceptrons, finite-size scaling, critical 
exponents, first-order phase transitions

\newpage

Some time ago W.Nadler and W.Fink \cite{NF} showed, for the model of the
perceptron, that
the transition from storable to unstorable pattern set sizes obeys
finite-size scaling (FSS) behavior. This transition is characterized
by the absence of an intrinsic length scale as is the correlation 
length for the usual phase transitions. 

Similar ``geometrical'' phase transitions without intrinsic length, 
are the satisfiability of random boolean expressions 
\cite{SK}, \cite{Zecchina}, the connectivity of random graphs \cite{Erdoes}, 
the quasispecies model of molecular evolution 
\cite{CF}, etc. All these models exhibit a sharp 
transitions for large values of the size of the corresponding system,
characterized by universal scaling functions, which describe the size-dependent
effects near the threshold.
Recently it was also shown that the statistical mechanics 
study of the K-SAT model is very  
useful for solving the hard computational 
NP-complete problem, as it represents,  since the nature of the
transitions which occur in it may help for the improvement of the 
efficiency of the search algorithms \cite{Nature}. 

In the present letter we study the FSS behavior of perceptrons in the
context of the usual FSS study known from different physical
systems \cite{Privman}-\cite{Binder}. For this purpose we use some of the
current definitions for shift and rounding of the transition, which occur
when the size of the system is finite.
 
The system, we are interested on, is a singe-layer perceptron storing a set 
of input
patterns $\{\xi_{i}^{\mu}\}, i=1,...,N; \mu=1,...,p$, drawn from a 
Gaussian distribution. By $N$ and $p$ we denote the numbers  of
the inputs and the patterns, respectively.

It is well known \cite{Cover}-\cite{Duda} 
that for the system  with one output unit,
Gaussian inputs and continuous couplings, the fraction of all 
the possible
input-output relations  of size $\alpha=p/N$ that 
can be
stored, called $P(\alpha,N)$, exhibits a smooth transition 
for finite value of $N$,
which becomes discontinuous
(step-like) at
$\alpha_c=2$ and in the thermodynamic limit
$N \rightarrow \infty$ (see Fig. 5.11 in ref\cite{Hertz} or  
Fig.3.4 in ref.\cite{Duda}).
The exact analytical expression  for
$P(\alpha,N)$ is \cite{Cover}:
\begin{equation}
\label{eq1}
P(\alpha,N) = 2^{1-p} \Sigma_{i=0}^{N-1}
(\begin{array}{c}p-1\\i\end{array}) .
\end{equation}
When the size of the system $N$ is large, eq.(\ref{eq1}) 
takes the asymptotic form:
\begin{equation}
\label{eq2}
P(\alpha,N) \approx \frac{1}{2} \left( 1 + Erf(\sqrt{\frac{N}{2 
\alpha}}
(2 - \alpha))\right)  ,
\end{equation}
revealing a FSS behavior with a scaling parameter
\begin{equation}
\label{eq3}
y = (\alpha - \alpha_c) N^{1/\nu}
\end{equation}
and a scaling exponent $\nu = 2$ near the transition point 
$\alpha_c=2$. 
The plot of $P(\alpha,N)$ in terms of the scaling parameter $y$
leads to the fall of all the curves with different $N$ 
onto a single one \cite{NF}. 

Because of the mean-field character of the model, the usual concept
of length and dimensionality become ambiguous. To avoid the lack of natural
geometric description in this case, one can choose the number of particles
(or the number of inputs in our case)
$N$ as a finite-size parameter and the dimensionality of the system can be
considered as arbitrary. Note that the
standard finite size scaling hypotheses, involving the notion of
correlation length, need a suitable extension for such systems \cite{Botet}.It
is easy to make
the analogy between the scaling parameter $y$ for the perceptron  
and the corresponding scaling
parameter for infinitely coordinated systems \cite{Botet}, defined by the
coherence number, which replaces the usual correlation length. 
The relation between the two models 
leads to the same scaling form, 
eq.(\ref{eq3}), expressed in
terms of the corresponding critical parameter.

Following the analogy with the conventional first-order transitions
\cite{Privman}-\cite{Binder}, we define as a transition point $\alpha_c(N)$
this value of the parameter $\alpha$, for which the derivative
$\left|\frac{\partial{P(\alpha,N)}}{\partial{\alpha}}\right|$ shows 
a maximum for large but finite values of $N$ 
($N$ being the size of the system) \cite{KN}.
This derivative becomes divergent in the thermodynamic
limit at $\alpha_c=2$. 

Using the above scheme and eq.(\ref{eq1}), we calculated\footnote{Here 
we would 
like to
mention that the time to find a solution diverges 
with $N$, reminding of the 
critical slowing down.}
the inflection
point of the function $P(\alpha,N)$ with respect to the parameter $\alpha$ for
different values of $N$  and we
identified the so-called shift exponent \cite{Privman}-\cite{Binder}: 
\begin{equation}
\label{eq4}
\alpha_c(N) - \alpha_c(\infty) \sim \frac{1}{N^\lambda} .
\end{equation}

We obtained the following dependence of the critical storage capacity
$\alpha_c(N)$:
\begin{equation}
\label{eq5}
\alpha_c(N) - \alpha_c(\infty) = \frac{e^{0.0328 - 
0.00006N}}{N^{1.00906}} ,
\end{equation}
for $N$ running between $N=8$ to $N=400$
and
\begin{equation}
\label{eq6}
\alpha_c(N) - \alpha_c(\infty) = \frac{e^{0.00571}}{N^{1.00104}} ,
\end{equation}
for $N$ between $N=100$ and $N=400$, Fig(1).

It becomes evident that for very large $N$, the following scaling 
dependence takes place:
\begin{equation}
\label{eq7}
\alpha_c(N) - \alpha_c(\infty) \sim \frac{1}{N} .
\end{equation}
Using the definition of the
shift exponent, eq.(\ref{eq4}), we obtain $\lambda = 1$.

\begin{figure}[t]
\begin{center}
\epsfysize=9cm
\leavevmode
\epsfbox[1 1 700 700]{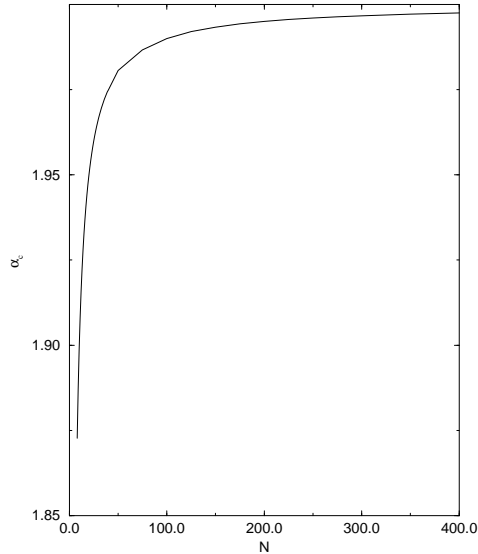}
\end{center}
\caption{
The critical storage capacity $\alpha_c(N)$ for values of $N$ within the
interval [8,400].}
\end{figure}

The other possibility to calculate $\lambda$ is by performing
analytical expansions for small values of the shift from 
the transition point $\alpha_c=2$,
using the asymptotic
expression eq.(\ref{eq2}). A straightforward calculations leads to the 
same result for the shift exponent, i.e., $\lambda=1$.
Note that the value of the shift exponent is obtained by 
defining the shift of the
transition point as the position of the maximum of 
$\left|\frac{\partial{P(\alpha,N)}}{\partial{\alpha}}\right|$, as is usually 
done in FSS analysis of physical systems \cite{Binder}. This value 
of $\lambda$ does
not coincide with the value $1/\nu=1/2$, as is usually expecting
from the analogy with the FSS in physical systems. As we see with the
present model, the coincidence of the values of the both exponents 
is not a necessary condition for the FSS to hold.

The finite shift of the critical point shows that the
perceptron belongs to the class of systems undergoing asymmetric
phase transitions. For symmetric phase transitions the shift is
zero and a typical example for such transition is the field-driven transition
in the Ising model, where the symmetry for the susceptibility $\chi(h,T,L) =
\chi(-h,T,L)$ takes place (here $h$ is the external field and $L$ is the finite
size of the system). Obviously, a finite shift in our case means that the 
above symmetry is broken, which is usually the case of transitions driven by 
temperature \cite{Binder}.

The result (\ref{eq7}) is similar to the well known result for
asymmetric temperature-driven first-order transitions in finite $d$-dimensional
systems with cubic symmetry, 
where the location of the shift by the maximal 
slope scales like $L^{-d}$ ($L$ being the finite size) 
\cite{Privman}-\cite{Binder}. Although our system is 
effectively finite (with size $N$) in one dimension,
we regard this analogy just as a formal one, 
because of the mean-field character of the interactions of the
present model and the absence of any boundary conditions imposed.

Apart of the definition of the transition point by the maximum 
of $P(\alpha,N)$, there is also another 
definition of the location of the transition for the
first-order transitions, which assumes the equilibrium of various 
phases near the point of the  
transition \cite{BK}. In contrast to the usual result
known for the  $q$-state Potts model \cite{Binder}, where the
shift is given by $L^{-d}$, $L$ being the size of the system and $d$ its
dimension, the definition used in ref.\cite{BK} leads to exponentially 
small corrections for the shift.
In the case of a perceptron, however, we can not make the close 
analogy following the
last scheme, because of the lake of various phases at equilibrium, which
is crucial for the application of this definition.

The two classes of transitions, symmetric and asymmetric, also show a
rounding behavior for $N$ finite, which is given by the scaling of the
width of the peak of the diverging observable. In other words, it is the 
interval over 
which the singularity is smeared out and which 
becomes increasingly sharp as the
finite dimension of the system goes to infinity.
In the concrete case of the simple perceptron this is the scaling of the width
of $\left|\frac{\partial{P(\alpha,N)|}}{\partial{\alpha}}\right|$, 
which determines
the rounding behavior. Using eq.(\ref{eq2}), the derivative reads:
\begin{equation}
\label{eq10}
\left|\frac{\partial{P(\alpha,N)}}{\partial{\alpha}}\right| \sim
\sqrt{\frac{N}{2\alpha}} \exp\{-\frac{[N (2-\alpha)^2]}{2\alpha}\} ,
\end{equation}
from where the scaling of the variance of the Gaussian distribution gives a
rounding exponent $\theta=\frac{1}{2}$. Note that a similar 
behavior with $N$ occurs for the shift and the variance of the 
generalization error in the case of a Bayesian perceptron with continuous
weights \cite{Gordon}. 

An interesting problem is what happens in the case of a  perceptron with binary
weights \cite{Krauth}. For this case the numerical analysis  for the
typical fraction shows a sharper behavior between the two regimes by 
increasing $N$,
but there is no definitive conclusion about the step-like behavior in the
thermodynamic limit \cite{Mertens}. An important 
investigation of the shift of the transition will be a similar calculation
of the position of the maximum of the above
derivative as we did for the continuous couplings model. 
This will probably lead to different results, since in the binary case 
the probabilities of separation as a function of $\alpha$ for different $N$
do not intersect at the same point.

In the general case
of a  perceptron, having a larger variety of outputs \cite{BHS},
and $p\geq d_{VC}$
($d_{VC}$ being the Vapnik-Chervonenkis (VC) 
dimension),
the fraction 
of all the possible
input-output relations obeys the following inequality \cite{Vapnik}:
\begin{equation}
\label{eq12}
P(\alpha,N) \leq 2^{1-p} \Sigma_{i=0}^{d_{VC}}(\begin{array}{c}
p-1\\i\end{array}) .
\end{equation}

It has been shown \cite{Opper} that in the thermodynamic limit
$N \rightarrow \infty$ ($p,d_{VC} \rightarrow \infty$) and keeping
$\alpha = \frac{p}{N}$ , and $\alpha_{VC} = \frac{d_{VC}}{N}$ fixed, 
the
VC-entropy shows different behavior above and below 
$\alpha=2\alpha_{VC}$,
which permits to relate the storage capacity of the network to its VC-
dimension via $\alpha_c \leq 2\alpha_{VC}$ 
$(\alpha_c\equiv\alpha_c(\infty))$ (In the case of a single layer
perceptron, treated at the beginning, $d_{VC} = N$, $\alpha_{VC} = 1$ 
and
$\alpha_c = 2$).

Eq.(\ref{eq12}) shows that for $N$-large, the
asymptotic form of the upper bound  $\bar{P}(\alpha,N)$ of the
fraction $P(\alpha,N)$ is given by:
\begin{equation}
\label{eq13}
P(\alpha,N)\leq \bar{P}(\alpha,N) = \frac{1}{2} \left[ 1 
+ Erf\left( \sqrt{\frac{N}{2\alpha}}
(2\alpha_{VC} - \alpha)\right)\right] ,
\end{equation}
leading to the same values
for the shift and the rounding exponents for the upper bound, 
known from the case of the simple perceptron.

Using the previous conclusion for the upper limit of $P(\alpha,N)$,
identifying $2\alpha_{VC}$ with some "upper" critical storage capacity,
and using the inequality between $\alpha_{VC}$ and $\alpha_c$, we
derive the following relations:
\begin{equation}
\label{eq14}
| \alpha(N) - \alpha_c| \geq |\alpha(N) - 2\alpha_{VC}| \sim 
\frac{1}{N^{\lambda}}
\end{equation}
with $\lambda=1$
and
\begin{equation}
\label{eq15}
|\alpha^{*}(N) - \alpha_c| \geq |\alpha^{*}(N) - 2\alpha_{VC}| \sim
\frac{1}{N^{\theta}}
\end{equation}
with $\theta=\frac{1}{2}$.

Taking into account the last expressions and the fact that the 
step-like behavior and the main characteristics of the upper bound
persist also in the general case by increasing the size of the system,  
we conclude that the shift and the rounding exponents for the upper bound, 
eq.(\ref{eq13}), are also upper bounds for the shift and 
rounding exponents in the general case \cite{Opper}.

In conclusion, in the case of a single-layer perceptron, using the analogy 
with the usual FSS theory, we derived the shift and the rounding critical 
exponents
$\lambda$ and $\theta$, respectively. 
Similar analysis in the general case gives results 
only for the upper limit of the fraction $P(\alpha,N)$. The full understanding
of the problem requires additional numerical simulations and an investigation
for every concrete case of architecture and machine.

\vskip0.5cm
{\bf Acknowledgments:}
E.K. warmly thanks for discussion the Workshop on
``Statistical Physics of Neural Networks'', organized by the
Max-Planck Institut-Dresden, 1999. 
This work is
supported by the Spanish DGES Project PB97-0076 and the Bulgarian Scientific
Foundation  Grant F-608.

\end{document}